\documentclass[3p, 11pt]{elsarticle}
\usepackage{graphicx,amscd,amsmath,amssymb,verbatim}
\usepackage{amsfonts,epsfig}
\usepackage{mathptmx}
\usepackage{setspace}
\usepackage{epsfig}
\usepackage{url}
\usepackage{multirow}
\usepackage{color}
\usepackage{subfigure}

\begin{document}

\journal{Elsevier}

\begin{frontmatter}

\title{When and how much the altruism impacts your privileged information? Proposing a new paradigm in game theory: The \textquotedblleft boxers game\textquotedblright}

\author{Roberto da Silva$^{1}$, Henrique A. Fernandes$^{2}$}

\address{$^1$Instituto de F{\'i}sica, Universidade Federal do Rio Grande do Sul, Av. Bento Gon{\c{c}}alves, 9500 - CEP 91501-970, Porto Alegre, Rio Grande do Sul, Brazil\\
$^2$Instituto de Ci{\^e}ncias Exatas, Universidade Federal de Goi{\'a}s, Regional Jata{\'i}, BR 364, km 192, 3800 - CEP 75801-615, Jata{\'i}, Goi{\'a}s, Brazil}

\begin{abstract}

In this work, we proposed a new $N$-person game in which the players can bet on two options, for example
represented by two boxers. Some of the players have privileged information about the boxers and part of them 
can provide this information to uninformed players. However, this information may be true if the informed player 
is altruist or false if he is selfish. So, in this game, the players are divided in three categories: informed and 
altruist players, informed and selfish players, and uninformed players. By considering the matchings 
($N/2$ distinct pairs of randomly chosen players) and that the payoff of the winning group follows
aspects captured from two important games, the public goods game and minority game, we showed quantitatively 
and qualitatively how the altruism can impact on the privileged information. We localized analytically the regions 
of positive payoffs which were corroborated by numerical simulations performed for all values of information and 
altruism densities given that we know the information level of the informed players. Finally, in an evolutionary version 
of the game ,we showed that the gain of the informed players can get worse if we adopted the following procedure: the 
players increase their investment for situations of positive payoffs, and decrease their investment when negative payoffs 
occur.

\end{abstract}

\end{frontmatter}

\setlength{\baselineskip}{0.7cm}

\section{Introduction}

Just over half a century ago, J. von Neumann and O. Morgenstern \cite%
{Neumann1944} probably had no idea of the extent that their work would have,
not only on applications in the economic behaviour but also in several areas
of biology, ecology, and mainly in the evolutionary theory of the species 
\cite{Maynard Smith 1982} culminating in the replicator dynamics \cite%
{Hofbauer2003,Nowak2006} and its branches (see, for example, the Refs. \cite%
{Hauert2005,Melbinger2010}). In fact, the game theory still persists as an
important theory of mathematical models which \textbf{studies} the choices
of optimal decisions under conflict or combat conditions. However, long
before the evolutionary aspects were studied in this context, the economical
applications were the primordial motivations of this theory, and leveraged
the construction of suitable concepts as Nash equilibrium, that contemplates
the existence of an equilibrium of mixed strategies in non-cooperative games 
\cite{Nash}. The idea was very simple but brilliant: in such equilibrium 
\textbf{condition}, when there are two or more players, no player will gain
by unilaterally changing its strategy. J. Nash was beyond and extended the
concept to cooperative games by performing a reduction to non-cooperative
games. His important works culminated in the Nobel Memorial Prize in
Economic Sciences jointly to R. Selten and J. Harsany \cite{Harsanyi and
Selten}.

An important concept in game theory arises when the game can be repeated
numerous times. This game is sometimes referred in literature as dynamic
game, or simply repeated game, in which the iteration effects can be
understood under two different theories: the deterministic and stochastic
game theories. One more interesting kind of dynamic game theory is one that
covers the stochastic case, and was introduced and formalized by Shapley in
1953 \cite{Shapley1953,Mertens1981,Solan2015}.

Looking in the other direction, games where the profit depends on the number
of investors have a direct similarity with the reality in the business
world. Here, the paradigmatic El farol bar problem (EFBP), which is a simple
model that shows how (selfish) players cooperate with each other in the
absence of communication, arises as an important starting point \cite%
{Arthur1994}. In that problem, a set of people must decide if they go to a
bar or stay at home. If they decide to go out and the bar is too crowded,
they probably will not have a funny night and stay at home would be more
interesting. They will have (or not have) a funny night according to a
threshold on the number of people in the bar. If less than a fraction (a
threshold) $x$ of the population decide to go to the bar, they will have a
better time than ones which decided to remain in their houses. On the other
hand, if more than a fraction $x$ of the population go to the bar, they will
have bad times when compared with the comfort of their homes.

Two physicists saw the potential of this idea, and a formalization of the
EFBP was given by Challet and Zhang \cite{Challet1997} in 1997, the
so-called Minority Game (MG) which was proposed as a rough model to describe
the price fluctuation of stock markets. In that game, an odd number of
players have to choose one of two options independently at each turn. The
players who end up on the minority side win the game. Following this line,
another multiple player game (strongly based on experimental economics of
the public goods), is the public goods game (PGG). In that game, multiple
players can contribute to a common fund with or without the presence of
communication in order to obtain benefits, and all money collected is
doubled, or triplicate, or simply corrected according to some multiplicative
factor. This game has been studied in many contexts by including optional
participation obtaining a fixed income by opting not to invest in the public
good, spatial diffusion effects and many other ingredients, both when
studying many public goods games \cite%
{Szabo2002,Hauert2003,Hauertetal2002,Pablo2017} and when considering only
one public good (see, for example, \cite%
{SilvaIJBC2010,SilvaBJP2008,SilvaLNE2006, SilvaPhysA2006}). Differently from
MG where the payoffs of players are inversely proportional to the number of
persons which choose the same group, in PGG the payoff depends on how many
persons invest in the fund.

The important feature of the MG is the absence of communication and noise.
However, in general, real situations have properties such as misinformation,
noise, selfishness, altruism, and some of these ingredients can coexist in
more complex scenarios. In this work, we propose a new game that captures
some aspects of both PGG and MG by taking into account some of those
properties as well as multiplayer games in order to show how the altruism
can affect the information of a group and change their payoffs. However, we
consider very different constraints, with the payoffs being determined
probabilistically and according to the information (and its propagation)
controlled only by some players. We call it the \textit{boxers game}, since
it was partly based on the idea of the payoffs obtained by gamblers in
boxing matches. However, the motivations can be easily extended to include,
for instance, horse-race or even gambling on the stock exchange, or even
more general situations.

The paper is organized as follows: In the next section, we formulate the
model. In Sec. \ref{Sec:Results}, we present our main results which are
separated in two parts. The first one considers the static, or fixed,
investments and the sampling of payoffs are analysed through Monte Carlo
(MC) simulations and analytical results. The second part of our results is
related to the dynamic investments in which a simple evolutionary aspect of
the game, based on reaction of the players according to their profit, is
studied via MC simulations. Finally, our summaries and conclusions are
presented in Sec. \ref{Sec:Conclusions}.

\section{The model}

\label{Sec:Model}

In this approach, we consider a $n$-person game in a scenario where each
player must choose between two boxers (groups): $A$ or $B$. Some players
(with density $\rho $) know that the boxer $A$ defeats the boxer $B$ with
probability $p>1/2$ (denoted as $A\succ B$) given, for example, their skills%
\textbf{. }Therefore, boxer $B$ defeats boxer $A$ (denoted as $A\prec B$)
with probability $1-p-q$ such that $q$ is the draw probability. Here, it is
important to mention that the information is incomplete since $p$ is not
exactly equal to 1 (in this case $q$ should be necessarily equal to $0$).
The misinformation, which has an important role in the payoff of the
players, was firstly formalized by the Nobel prize laureate in Economic
Sciences, John C. Harsanyi \cite{Harsanyi}. However, we propose a different
approach: by considering the similarities with a boxing match, we establish
that the payoff of the player in a group $A$ or $B$ depends on some features
considering some protocols that remember both PGG and MG. So, we set up
three important fundamentals that govern our game:

\begin{enumerate}
\item MG protocol: in the case of victory, the payoff of the players
(winners) depends on how many persons have bet in the losing group;

\item PGG protocol: the profit obtained for the group must be divided
equally among the players of that group;

\item The gain is proportional to the invested amount.
\end{enumerate}

By denoting $m_{A}^{(i)}$ as the payoff of the $i$-th player that has chosen
the group $A$, which, without loss of generality, is the group where a
number of people have privileged information ($A\succ B$ with probability $%
p>1/2$), we have 
\begin{equation}
m_{A}^{(i)}=\left\{ 
\begin{array}{ll}
s_{i}^{(A)}\frac{\sum_{i=1}^{n_{B}}s_{i}^{(B)}}{\sum_{i=1}^{n_{A}}s_{i}^{(A)}%
} & \text{with probability }p \\ 
-s_{i}^{(A)} & \text{with probability }1-p-q \\ 
0 & \text{elsewhere}%
\end{array}%
\right.  \label{Eq: profit-A}
\end{equation}
and naturally 
\begin{equation}
m_{B}^{(i)}=\left\{ 
\begin{array}{ll}
s_{i}^{(B)}\frac{\sum_{i=1}^{n_{A}}s_{i}^{(A)}}{\sum_{i=1}^{n_{B}}s_{i}^{(B)}%
} & \text{with probability }1-p-q \\ 
-s_{i}^{(B)} & \text{with probability }p \\ 
0 & \text{elsewhere}%
\end{array}%
\right.  \label{Eq: profit-B}
\end{equation}

The reader should observe that the return, in case of a win, is the sum of
the investment of all players in the loosing group multiplied by a factor
that distributes equally the gain between the participants of the winning
group: $\alpha _{A,B}=s_{i}^{(A,B)}/\sum_{i=1}^{n_{A,B}}s_{i}^{(A,B)}$. It
is important to mention that for equal investments, $s_{i}^{(A,B)}=s$, for
all players, the payoff for each player is $\alpha _{A,B}=1/n_{A,B}$.

This game is a good example of multiple zero-sum game, since the sum of
payoff of the players in a match is 0, which means that if one or more
players gain a certain amount, the other ones lose this same amout, i.e., 
\begin{equation}
m^{(i)}=-\sum_{j\neq i}m^{(j)}  \label{Eq:sum_zero}
\end{equation}

As a partition and the sequences of \textquotedblleft
bets\textquotedblright\ are given, respectively, by $\{n_{A},n_{B}\}$, $%
s_{1}^{(A)},...,s_{n_{A}}^{(A)}$ and $s_{1}^{(B)},...,s_{n_{B}}^{(B)}$, a
simple calculation can be performed to provide us the expected payoff of the 
$i-$th player: 
\begin{equation}
E\left[ m_{A}^{(i)}\right] =ps_{i}^{(A)}\frac{\sum_{i=1}^{n_{B}}s_{i}^{(B)}}{%
\sum_{i=1}^{n_{A}}s_{i}^{(A)}}-(1-p-q)s_{i}^{(A)}  \label{Eq:expect_A}
\end{equation}%
and 
\begin{equation}
E\left[ m_{B}^{(i)}\right] =(1-p-q)s_{i}^{(B)}\frac{%
\sum_{i=1}^{n_{A}}s_{i}^{(A)}}{\sum_{i=1}^{n_{B}}s_{i}^{(B)}}-ps_{i}^{(B)}
\label{Eq:expect_B}
\end{equation}

In this regard, it is important to think of a mathematical point of view
without taking into consideration the complex interaction that exists among
players and how some of them ($n\rho $) can use their privileged information
to make money. The problem is: giving the partition $\{n_{A},n_{B}\}$ and
the bets $\left\{ s_{i}^{(A)}\right\} _{i=1}^{n_{A}}$ and $\left\{
s_{i}^{(B)}\right\} _{i=1}^{n_{B}}$ conditioned to this partition, how to
compute the average payoff?

For example, if, by hypothesis, we know the probability distribution $\Pr
(n_{A,B})$ that determines how many individuals choose the group $A(B)$ as
well as the conditional $f(s|A,B)$ probability distribution function (pdf)
that governs the probability of a player which invests in the group $A(B)$
to perform a proposal between $s$ and $s+ds$, we would have the average
payoffs of the individuals in the groups $A$ and $B$ given, respectively, by 
\begin{eqnarray}
\left\langle \left\langle m_{A}\right\rangle \right\rangle &=&p\left\langle
s^{(B)}\right\rangle \frac{\overline{n_{B}}}{\overline{n_{A}}}%
-(1-p-q)\left\langle s^{(A)}\right\rangle  \label{Eq:averages} \\
&&  \notag \\
\left\langle \left\langle m_{B}\right\rangle \right\rangle
&=&(1-p-q)\left\langle s^{(A)}\right\rangle \frac{\overline{n_{A}}}{%
\overline{n_{B}}}-p\left\langle s^{(B)}\right\rangle  \label{Eq:averages3}
\end{eqnarray}%
where $\overline{n_{A,B}}=\sum_{n_{A,B}=0}^{n}n_{A,B}\Pr (n_{A,B})$ and $%
\left\langle s\right\rangle _{A,B}=\int_{0}^{\infty }sf(s|A,B)ds$.

At this point, the interactions among players, misinformation, and other
relevant features, should lead to more exciting studies for this game since
no indication produces the following distributions: $\Pr (n_{A,B})$ and $%
f(s|A,B)$. So, once we showed that our very different approach avoids the
Bayesian formalism of the incomplete information game theory \cite{Harsanyi}%
, now we are able to define the dynamics for this complex scenario by
presenting how the players propagate their privileged information. First, in
real situations the information can be propagated with good or bad
intentions. Among the players that know the bias in a group (there are $%
n_{I}=\rho n$ players with privileged information), a fraction $\alpha $ of
them propagate this information with good intention (i.e., saying the
truth). On the other hand, there are $(1-\alpha )\rho n$ players which only
wish to maximize their profit (sincerity is not their main feature) and will
influence all the other players to invest in the other group. In addition, $%
(1-\rho )n$ players do not know the chances of each group (or of the
corresponding boxer) and therefore, depend on the information from the other
players.

So, we have a model with three types of players:

\begin{enumerate}
\item \textbf{IA} -- (Informed and altruist players) -- They know the
probability $p>1/2$ of a group winning and propagate this information;

\item \textbf{IS} -- (Informed and selfish players) -- They know the
probability $p>1/2$ of a group winning but suggests to the misinformed
players to invest their contribution in the opposing group lying to them;

\item \textbf{U} -- (uninformed players) -- They have no information and
depend on the information given by the two first groups.
\end{enumerate}

In addition, the model has four parameters: $p$, $q$, $\rho $, and $\alpha $%
, and the profit in each round is given by Eqs. (\ref{Eq: profit-A}) and (%
\ref{Eq: profit-B}).

Now let us define the dynamics of the game by supposing (without loss of
generality) that $p>1/2$ is the probability of the group $A$ defeats the
group $B$, as shown in Table \ref{encounters}.

\begin{table}[tbp] \centering%
\begin{tabular}{ll}
\hline\hline
\textbf{Players} & \textbf{Decision} \\ \hline
I versus I (A or S) & Both players go to the group A \\ 
&  \\ 
U versus U & Each player choose a group \\ 
& with probability 1/2 \\ 
&  \\ 
IA \ versus U & Both players go to the group A \\ 
&  \\ 
IS versus U & The IS player goes to the group A \\ 
& and the U player goes to the group B \\ \hline\hline
\end{tabular}%
\caption{Possible decisions according to the different encounters among the
players\label{encounters}}%
\end{table}%

The group $A$ is composed by part of the informed players as well as a
fraction of uninformed ones that are convinced by the informed altruist
players to invest in this group and half of uninformed players which
interact with other uninformed players. On the other hand, the group $B$
consists of uninformed players convinced by the informed selfish ones to
invest in it and the other half of uninformed players which find other
uninformed players.

\section{Results}

\label{Sec:Results}

Now we present our main results. In the next subsection we present the most
of our results, which is an exploration of the main properties of the game
considering that all players invest the same quantity along the time: $%
s_{1}=s_{2}=...=s_{n}=s$, which remains fixed over the different iterations
of the game. In our numerical approach, we consider in total $n=1000$
different players and s is made equal to 1. In the following subsection, we
consider an interesting evolutionary aspect of the game, based on reactions
of the players according to their gain or loss, so the values $s_{i}(t)$,
evolve over time but $s_{i}(t=0)=1$, for all players: $i=1,2,...,n$.

\subsection{Results I: Sampling and probability theory}

We first consider a more natural situation where all players invest the same
quantity, $s_{1}=s_{2}=...=s_{n}=s$ in every iteration of the game. In this
case, we have 
\begin{eqnarray*}
\left\langle \left\langle m_{A}\right\rangle \right\rangle &=&\frac{%
\overline{n_{B}}}{\overline{n_{A}}}p-(1-p-q) \\
&& \\
\left\langle \left\langle m_{B}\right\rangle \right\rangle &=&(1-p-q)\frac{%
\overline{n_{A}}}{\overline{n_{B}}}-p
\end{eqnarray*}

According to Table \ref{encounters}, along with the fraction of players, we
have 
\begin{eqnarray*}
\overline{n_{A}} &=&N\rho +(1-\rho )\alpha \rho N+\frac{1}{2}(1-\rho )^{2}N
\\
&& \\
\overline{n_{B}} &=&(1-\rho )(1-\alpha )\rho N+\frac{1}{2}(1-\rho )^{2}N
\end{eqnarray*}%
where $\rho $ is the density of informed players, $\alpha $ is the density
of altruist informed players, and $\overline{n_{A}}+\overline{n_{B}}=N$,
i.e., the number of players is held constant.

The rate $\overline{n_{A}}/\overline{n_{B}}$ yields 
\begin{equation*}
\frac{\overline{n_{A}}}{\overline{n_{B}}}=f(\alpha ,\rho )=\frac{%
\rho+(1-\rho )\alpha \rho +\frac{1}{2}(1-\rho )^{2}}{(1-\rho )(1-\alpha
)\rho +\frac{1}{2}(1-\rho )^{2}}
\end{equation*}

Thus we can write 
\begin{equation}
\left\langle \left\langle m_{A}\right\rangle \right\rangle=\frac{%
(1-\rho)(1-\alpha )\rho +\frac{1}{2}(1-\rho )^{2}}{\rho +(1-\rho )\alpha
\rho +\frac{1}{2}(1-\rho )^{2}}p-(1-p-q)  \label{Eq:payoffa}
\end{equation}
and 
\begin{equation}
\left\langle \left\langle m_{B}\right\rangle \right\rangle =\frac{%
\rho+(1-\rho )\alpha \rho +\frac{1}{2}(1-\rho )^{2}}{(1-\rho )(1-\alpha
)\rho +\frac{1}{2}(1-\rho )^{2}}(1-p-q)-p  \label{Eq:payoffb}
\end{equation}

It is interesting to observe that for $\left\langle \left\langle
m_{A}\right\rangle \right\rangle >0$, i.e., the informed players have profit
in average. This situation leads to 
\begin{equation}
p>p_{c}=(1-q)\frac{\rho +(1-\rho )\alpha \rho +\frac{1}{2}(1-\rho )^{2}}{%
\rho +(1-\rho )^{2}+(1-\rho )}  \label{Eq:critical_information}
\end{equation}%
This means that $p_{c}$ is the minimum information level necessary to obtain
profit. Another alternative is to think of the level of information $p$
(probability of the favourite group to win) and the density of informed
players $\rho $. As we have them in hand, the altruism level required to
obtain profit is then given by 
\begin{equation}
\alpha >\alpha _{c}=\frac{2p-(1-q)(1+\rho ^{2})}{2(1-q)\rho (1-\rho )}
\label{Eq:critical_altruism}
\end{equation}

We can consider applications where $q>0$, and this deserves a future
investigation. But for the sake of simplicity, by thinking in boxers, a draw
is very rare, thus we make $q=0$ for all the results obtained from here in
this work.  

In this work, we perform numerical simulations based on turns. One turn is
averaged over $N_{run}$ different time series corresponding to independent
runs. Each time series corresponds to a sequence of iterations (rounds). In
each round, $N/2$ pairs of players are randomly matched and all players
necessarily participate once (such as performing a matching in a graph). In
each round, the players take a decision according to the information and
nature of their partners determined by the Table \ref{encounters}. So, for
each turn $t$, the average payoff of bettors in the group $A(B)$, calculated
via MC simulations, is obtained by a general formulae: 
\begin{equation*}
payoff_{A(B)}(t)=\frac{1}{n_{A(B)}N_{run}}\sum_{i=1}^{n_{A(B)}}\sum%
\limits_{j=1}^{N_{run}}payoff_{A(B)}(i,j,t)
\end{equation*}%
For fixed values of $t$ and $j$, one has that $payoff_{A(B)}(i,j,t)$ is the
same for all $i=1,2,3...,n_{A(B)}$, i.e., there is no difference among the
payoffs of the different players in same group, in the same run $(j)$ of the
same turn $(t)$, since the investments in this version of the game are fixed
and made equal to 1 for all players during all the iterations. However, as
shown in the next subsection, this is not the truth when the reaction of the
players to their profit is taken into consideration.

\begin{figure}[h]
\begin{center}
\includegraphics[width=1.0\columnwidth]{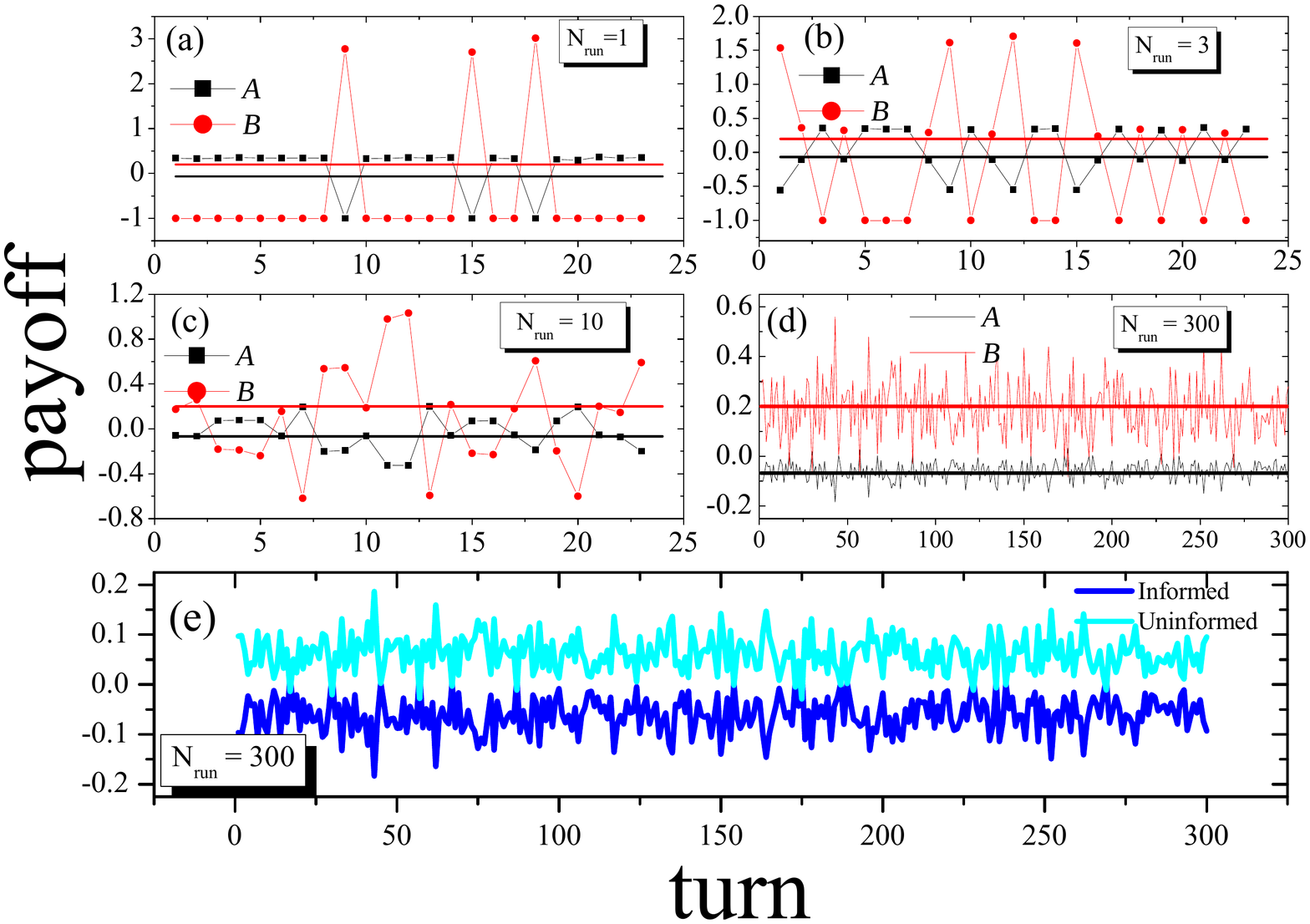}
\end{center}
\caption{Payoffs for different turns of the groups $A$ (black points) and $B$
(red points) for a given set of parameters. The straight horizontal lines
(average results) correspond to our theoretical predictions, Eqs. (\protect
\ref{Eq:payoffa}) and (\protect\ref{Eq:payoffb}). We analyze turns composed
by different number of runs, $N_{run}$ as can be seen in the plots (a), (b),
(c), and (d). The noisy results around these lines correspond to the MC
simulations. A good agreement can be observed for $N_{run}=300$ runs. Plot
(e) shows the payoffs for different turns of the informed players (which is
equal to the payoffs of all players in A) (dark blue line) and also shows
the payoffs of the uninformed players when they are not conditioned only to
the group $B$ (light blue line), i.e., the payoff is averaged over all
uninformed players. This last one is smaller than the case where uninformed
players are averaged conditioned to the group $B$ (see plot (d)).}
\label{Fig:MCXtheory}
\end{figure}

Figure \ref{Fig:MCXtheory} shows the payoffs obtained by MC simulations.
These estimates are compared with our theoretical result predicted by Eqs. (%
\ref{Eq:payoffa}) and (\ref{Eq:payoffb}). In Figs. \ref{Fig:MCXtheory} (a),
(b), (c), and (d), we show the results for the average payoff of players
that have bet in $A$ and $B$, for $N_{run}=1$, 3, 10, and 300, respectively.
We can observe a good agreement between our numerical and theoretical
predictions.

We complete our analysis with the plot shown in Fig. \ref{Fig:MCXtheory} (e)
which shows the average payoff of the informed and uniformed players. It is
important to observe that all informed players bet in the group (or boxer) $%
A $, but not all player that bets in $A$ is an informed player according to
our rules. On the other hand every, player that bets in the group $B$ is
necessarily an uninformed player. Of course, we do not observe difference
between the average payoff of the group formed by bettors in the group $A$
and the group formed by informed players, however we have that average
payoff of the uniformed players group even being greater than that of the
informed players group, is smaller than the group formed by the uninformed
players that bet in $B$ (average over uninformed players conditioned to the
group $B$). This shows that uninformed players which goes to the group $A$
are the responsible for this decrease or for getting worse the average of
the uninformed players from a general point of view.

By looking into the Fig. \ref{Fig:MCXtheory} (a), (b), (c), and (d), it
turns interesting to understand the influence of $N_{run}$ in our results,
or, in other words, we are wondering how the central limit theorem is
working here. When considering small values of $N_{run}$, we have that the
payoff switch between two different values. As $N_{run}$ enlarges, the
superposition among different runs generates a continuous distribution of
payoffs which we expect to be a gaussian distribution of payoffs over the
different turns. 
\begin{figure}[h]
\begin{center}
\includegraphics[width=1.0\columnwidth]{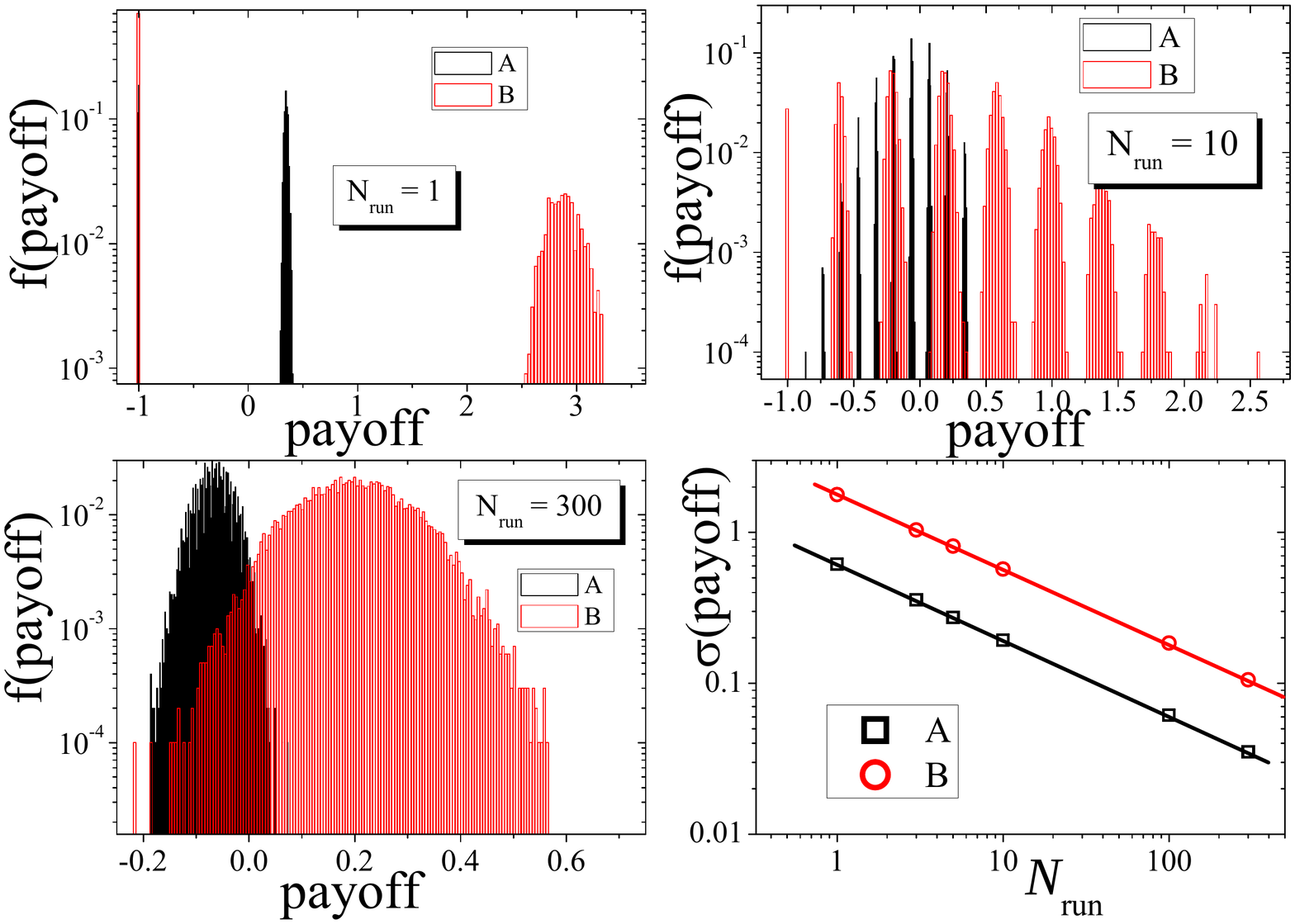}
\end{center}
\caption{Plots (a), (b), and (c) show the distribution of payoff for
different values of $N_{run}$ for the bettors in the groups $A$ and $B$. We
can observe a mixing distribution composed by a gaussian peak and a delta
peak which migrates to a unique gaussian distribution in both cases for
bettors in $A$ and $B$ (in mono-log scale). Plot (d) shows the variance of
payoff over different turns as function of $N_{run}$ by illustrating the
expected fitting $\protect\sigma (payoff)\sim N_{run}^{-1/2}$.}
\label{Fig_distributions}
\end{figure}

Figure \ref{Fig_distributions} shows the distribution of payoff over the
turns. The plots (a), (b), and (c) show the payoff frequencies for different
values of $N_{run}$ for the bettors in the groups $A$ and $B$. We can
observe a mixing distribution composed by a gaussian peak and a delta peak
migrating to a unique gaussian distribution in both cases: for bettors in $A$
and $B$ (in mono-log scale) as $N_{run}$ enlarges. The mix distribution
occurs because for $N_{run}=1$, each player loses or gains one unit with his
group dividing the quantity that changes along different rounds, generating
the Gaussian distribution. However it is interesting to observe that the
convergence occurs migrating from this bi-modal distribution, passing by a
multi-modal distribution (see for example $N_{run}=10$) until that such
modes are fused in a unique gaussian distribution. The central limit theorem
can also be observed by considering the standard deviations over the
different turns, calculated as: 
\begin{equation}
\sigma (payoff_{A(B)})=\sqrt{\frac{1}{(N_{turns}-1)}\sum_{t=1}^{N_{turns}}%
\left( payoff(t)-\overline{payoff}\right) ^{2}}\text{,}
\label{Eq:var_payoff}
\end{equation}%
where it is expected that $\sigma (payoff_{A(B)})\sim N_{run}^{-1/2}$. As
can be seen in Fig. \ref{Fig_distributions} (d), our results are in complete
agreement with each other. With this result, we finish our analysis about
sampling and agreement between MC simulations and theory.

\begin{figure}[h]
\begin{center}
\includegraphics[width=1.0\columnwidth]{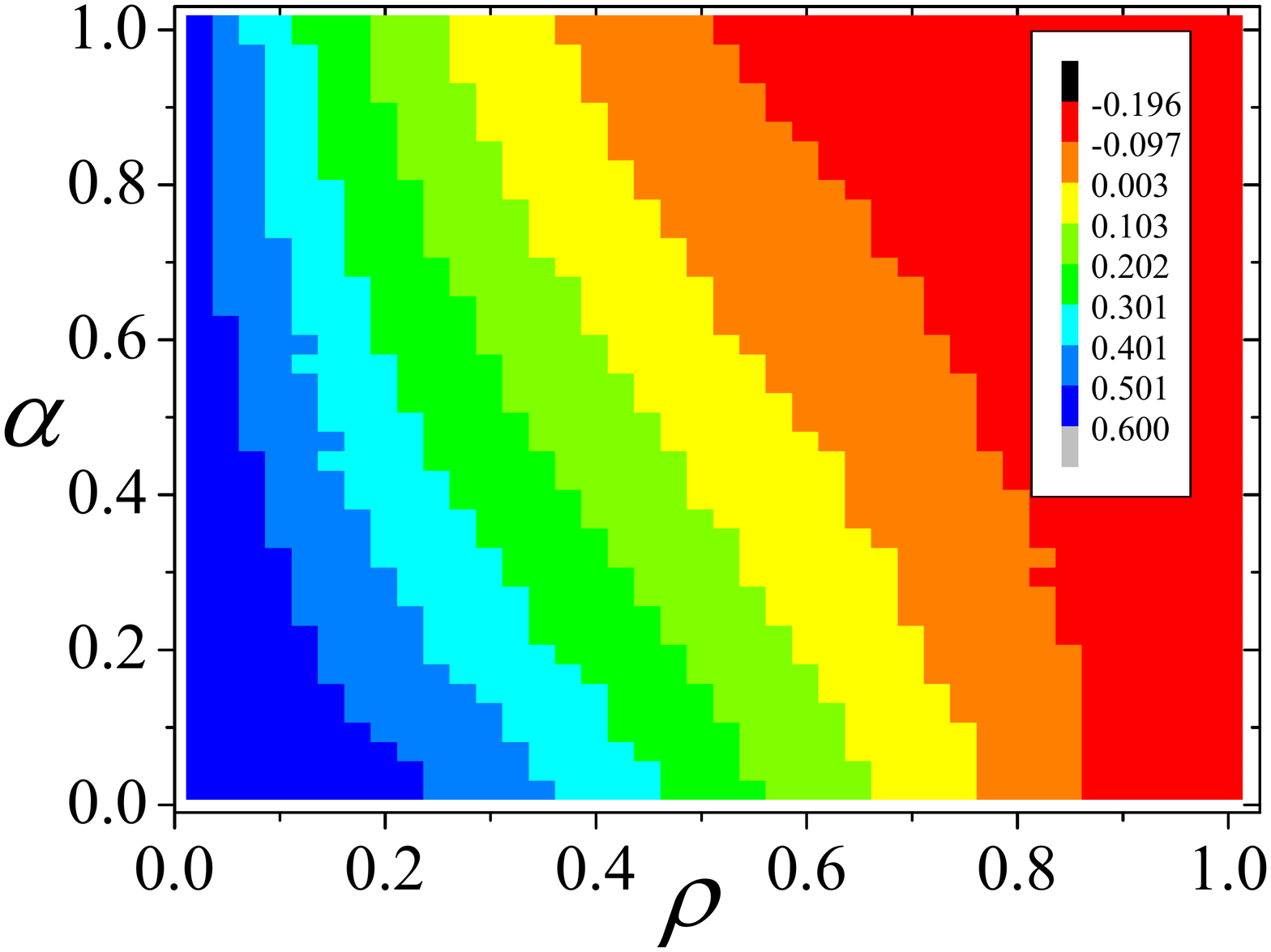}
\end{center}
\caption{MC simulations for the payoff of the group $A$ (group of the
informed players) by considering all possible values of information and
altruism for $p=0.8$.}
\label{Fig:payxalfaxrho}
\end{figure}

\begin{figure}[h]
\begin{center}
\includegraphics[width=1.0\columnwidth]{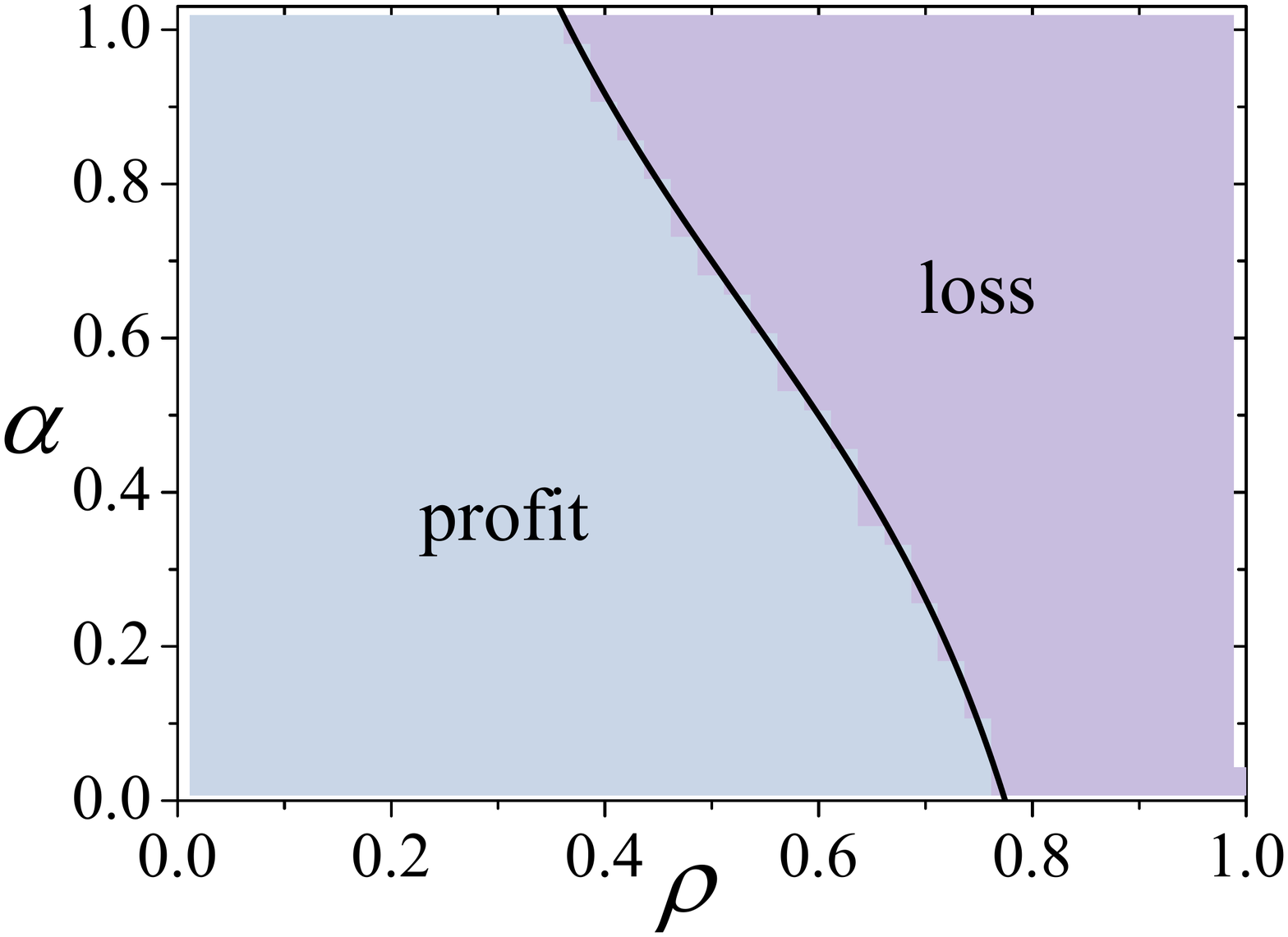}
\end{center}
\caption{Payoff diagram for $p=0.8$. The diagram was obtained from MC
simulations while the continuous curve shows the theoretical result
predicted by Eq. (\protect\ref{Eq:critical_altruism}). Below the curve we
have a positive payoff whereas above it we have a negative payoff.}
\label{Fig:payoff_diagram}
\end{figure}

Now, we explore the relationship between altruism and information in this
game. Since the altruism influences the information, we are wondering how
the payoff vary for all possible values of information, $\rho $, and
altruism, $\alpha $ for a fixed value of $p$, for instance, $p=0.8$. Figure %
\ref{Fig:payxalfaxrho} shows the payoff of players of the group $A$ whereas
this is the group of informed players. We can observe iso-payoff curves
which depend on values of $\rho $ and $\alpha $. 
From this figure, one can observe that for low densities of informed players
there exist only altruist players with positive payoff since this is a
profitable scenario, i.e., high-level information: $p=0.8$. However, for
high density of informed players, the pure altruism can bring strong damages
for the group and the information is not a trump for the game. In this case,
it is interesting to obtain the minimal altruism level required to reach a
positive payoff in the scenario predicted by Eq. (\ref{Eq:critical_altruism}%
). In Fig. \ref{Fig:payoff_diagram}, we show the diagram for all possible
values of $\rho $ and $\alpha $ for $p=0.8$. The positive (gray) and
negative (purple) payoff regions predicted by MC simulations are regions
where each set $(\rho ,\alpha )$ yields a profit or loss for the players,
respectively. The continuous curve (in black) shows the theoretical
prediction of the threshold between the profit and loss obtained by Eq. (\ref%
{Eq:critical_altruism}). 

\begin{figure}[h]
\begin{center}
\includegraphics[width=1.0\columnwidth]{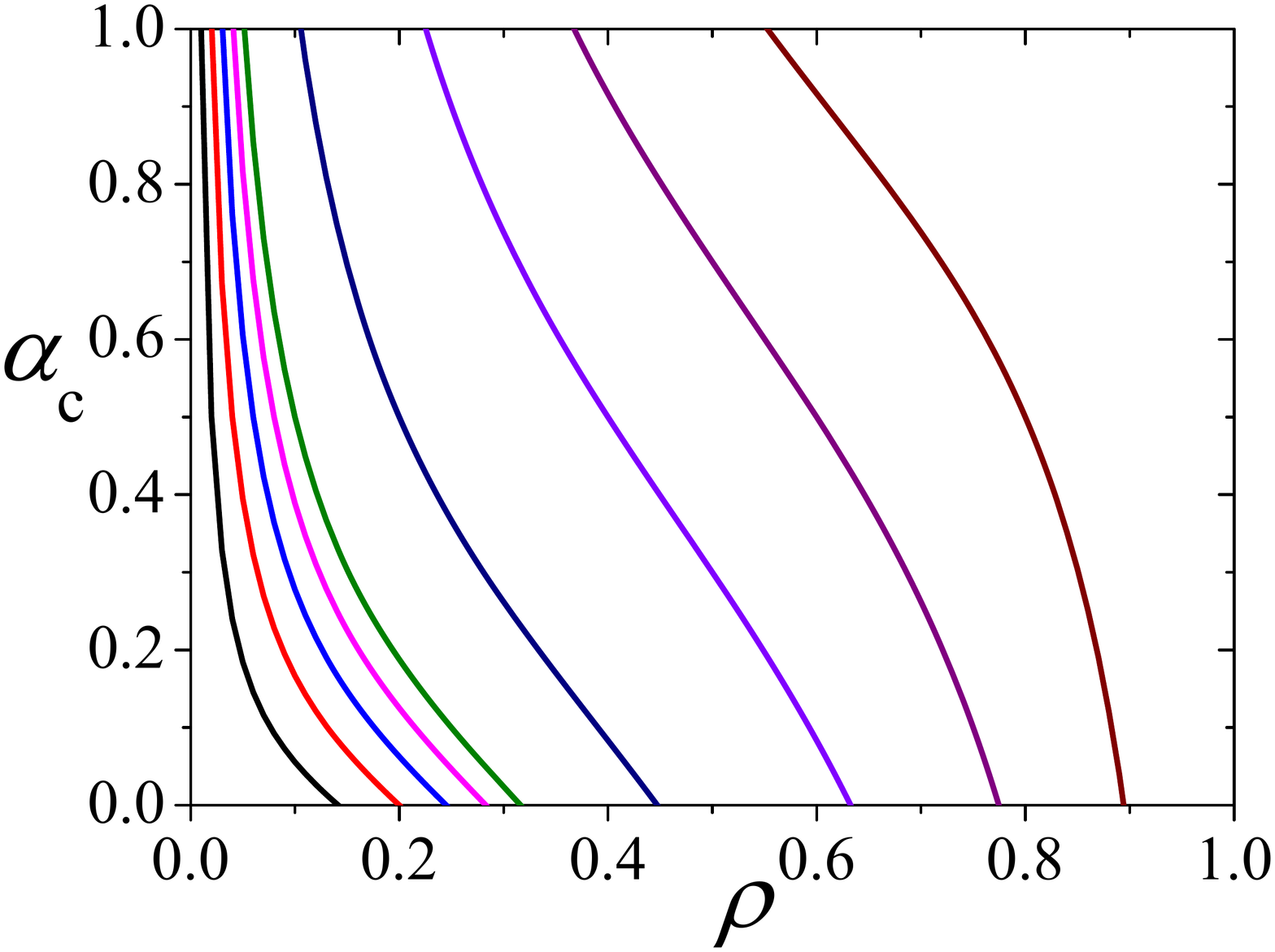}
\end{center}
\caption{Plots of $\protect\alpha _{c}$ $\times \protect\rho $, for
different values of $p$, predicted by Eq. (\protect\ref{Eq:critical_altruism}%
). From the left to the right, we show curves which correspond to the
following values of $p$: 0.51, 0.52, 0.53, 0.54, 0.55, 0.60, 0.70, 0.80, and
0.90.}
\label{Fig:alfacxrho}
\end{figure}

One can observe a perfect agreement between the simulations and our
analytical approach. Finally, we focus our attention on the influence of the
information level ($p$) on the transition curves $\alpha _{c}$ versus $\rho $
according to Eq. (\ref{Eq:critical_altruism}). Figure \ref{Fig:alfacxrho}
shows $\alpha _{c}$ $\times \rho $ for different values of $p$. As $p$
increases, we observe the evolution of the curves $\alpha _{c}\times \rho $.
It is interesting to observe that for $\rho >\rho _{c}=\left( 2p-1\right)
^{1/2}$ (this value is obtained by making $\alpha _{c}=0$ in Eq. (\ref%
{Eq:critical_altruism}), even for a population entirely formed by
non-altruist players, the payoff is always negative. So, our work shows that
in a population with people that possess some privileged information, the
payoff of the informed players is deeply changed by the altruism level. In
addition, there is a critical altruism level which separates the profit from
the loss in the payoff of the players. Our analytical results are in
complete agreement with MC simulations. 

\subsection{Some evolutionary aspects: reactive boxers game}

Here, we study a simple variation of the game where players may change their
investment according to their results. The evolution is not exactly as that
one conventionally considered in game theory where the strategies are copied
according to, for instance, the best payoff in the exact sense of Darwinian
evolutionary theory. In the right case of this paper, we propose a simple
evolutionary dynamics where the player only tries to adjust his investment
according to his personal losses or gains from a similar way as that used in
the ultimatum game studied in Ref. \cite{rdasilva2016} or in the public
goods game (see for example \cite{SilvaIJBC2010,SilvaLNE2006,SilvaPhysA2006}%
). Basically, one considers a parameter $\varepsilon >0$ used to allow the
players to make small adjustments depending on their payoff. If a player $i$
has invested $s_{i}(j)$ at turn $j$ of a certain run, and his payoff is
positive, then his investment is increased in the next round: $%
s_{i}(j+1)=s_{i}(j)+\varepsilon $. However, if the payoff is negative the
investment is decreased in the next round: $s_{i}(j+1)=s_{i}(j)-\varepsilon $%
. So, in this approach, we are able to analyse how the dynamical evolution
of the investments affects the wealth of the players.

For this particular analysis it is convenient to consider the cumulative
wealth defined as 
\begin{equation*}
W(t)=\sum\limits_{\tau =1}^{t}payoff(\tau )
\end{equation*}%
and, mainly, the average gain per turn: $g(t)=\frac{1}{t}W(t)$ which
measures the real average gain.

We add these ingredients to our previous study and repeat the MC simulations
to calculate $g(t)$ and its standard deviation $\sigma (g)(t)$ by
considering the different gains among the players in each group. It is
important to observe that when compared with the standard deviation
calculated in Eq. \ref{Eq:var_payoff}, this measures the variation over the
players (internal) while the first measures the variation over the turns
(external).

\begin{figure}[h]
\begin{center}
\includegraphics[width=1.0\columnwidth]{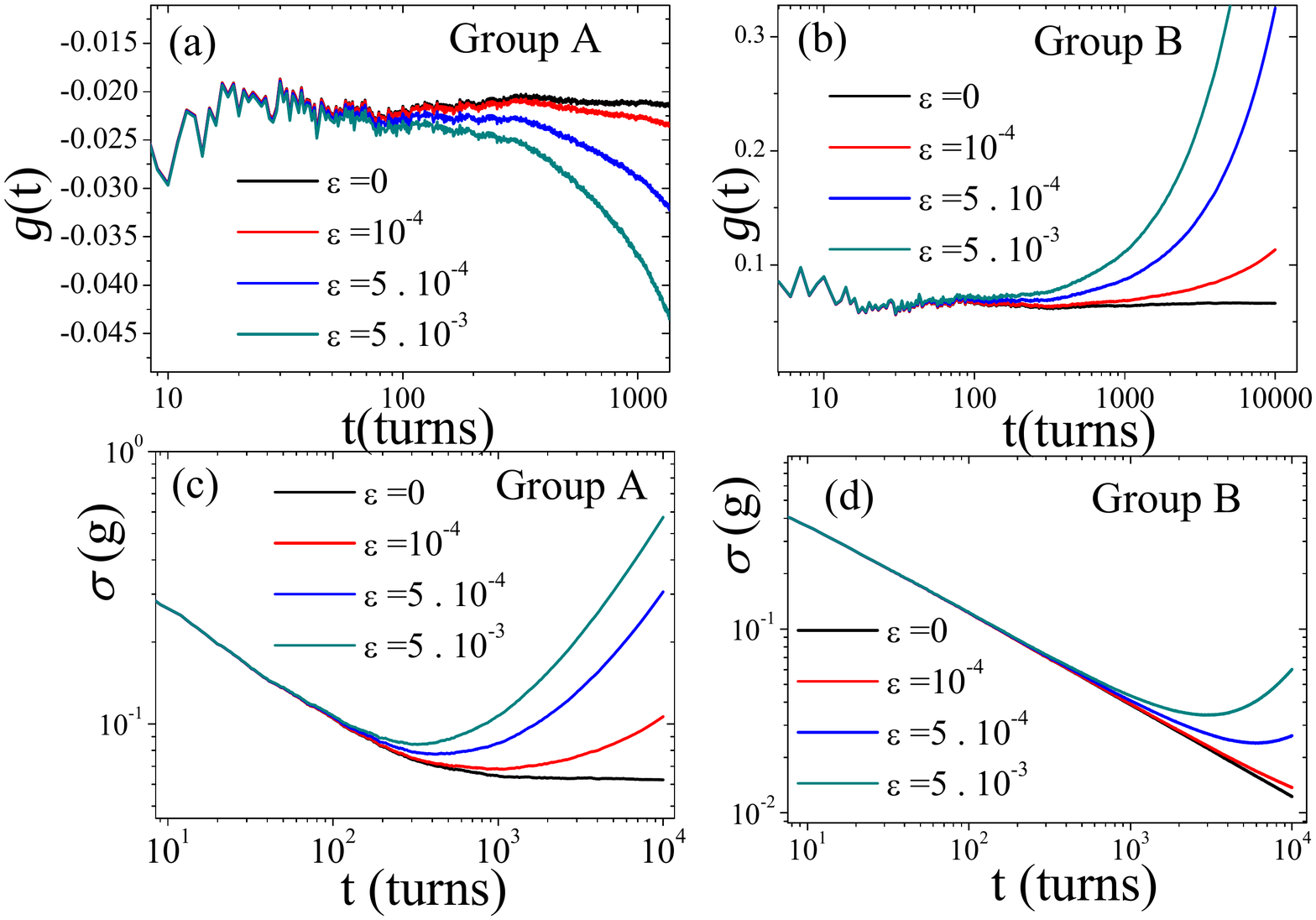}
\end{center}
\caption{Effects of the evolutionary investment on the gains of the players.
Plots (a) and (b) shows the gains of the groups $A$ and $B$, respectively,
for $p=0.7$, $\protect\rho =1/2$, and $\protect\alpha =1/2$. Plots (c) and
(d) correspond, respectively, to standard deviations of the gains of the
groups $A$ and $B$. All these quantities are average over $N_{run}=300$
runs. }
\label{Fig:Evolutionary}
\end{figure}

The variance $\sigma (g)\ $is also averaged over $N_{run}=300$ time series.
Here, we also choose the same bad situation for the group $A$ ($p=0.7$, $%
\rho =1/2$, and $\alpha =1/2$) and start the simulations with all players in
the same initial unit investment ($s_{i}(0)=1$, for $i=1,2,...n$). The
results for the different groups are presented in Fig. \ref{Fig:Evolutionary}%
. For $\varepsilon =0$, the gains remain approximately the same as presented
in the beginning of previous subsection since they correspond to similar
samplings. However for $\varepsilon >0$ the situation is disastrous for 
bettors in the group $A$ which is opposite for the bettors in the group $B$
that have a considerable increase in average as shown respectively in Figs \ref%
{Fig:Evolutionary} (a) and (b). 

It is important to observe that the variance in the bettors in the group $A$
is due to the difference between informed and uninformed ones since the
informed bettors necessarily have the same gain at the same turn. However,
this does not occur for the bettors in the group $B$, in which the variance
is calculated over many different wealths. This is clear in Figs. \ref%
{Fig:Evolutionary} (c) and (d) presented in $\log -\log $ scale. We can
observe that the variance $\varepsilon =0$ for the group $B$ has an
algebraic decay $\sigma (g)(t)\sim t^{-1/2}$, indicating a diffusive effect,
since it means $\sigma (W)(t)\sim t^{1/2}$. However, we do not observe this
same behaviour for the group $A$, which does not present a power law for $%
\varepsilon =0$. For different values of $\varepsilon $, the standard
deviation first decreases and then, after some time, enlarges which
indicates the corroboration with the anomalous behaviour of time evolution of
gain.

We also analyse other prosperous situations for the group $A\ $for $%
\varepsilon =0$, according to diagrams obtained in the last subsection, but
the effects of the evolutionary dynamics also destroy their profit, showing
that evolutionary dynamics is not good to the informed players differently than 
situation represented for the group $B\ $.  

\section{Summaries and brief conclusions}

\label{Sec:Conclusions}

In conclusion, we proposed a new game which presents an alternative social
paradigm in economic scenarios with incomplete information. Our results
claim that benefits of information can be destroyed under high altruism,
since the informed players, even with high level information (high $p$), can
gain more times but small quantities and the defeats, even rare, are
rentless. On the other hand, even earning few times, the uninformed players
can outweigh their losses, which are in average more frequent, but
inexpressive in absolute value.

In an evolutionary version of the game, we show that the gain of the
informed players can get worse if the following approach is adopted: the
player increases its investment for positive payoffs, and decreases the
investment for negative payoffs.

The results of our analytical approach are corroborated by MC simulations,
and the findings suggest that a lot of new applications may be considered in
future investigations, including, for instance, the study of spacial effects
through the analysis of this game in different networks or even considering
the evolutionary aspects on the strategies.

\textbf{Acknowledgements --} The authors would like to thank CNPq (National
Council for Scientific and Technological Development) for the partial
financial support.

\bigskip

\bigskip


\begin{thebibliography}{99}
\bibitem{Neumann1944} J. von Neumann e O. Morgenstern, Theory of Games and
Economic Behavior, Princeton University Press (1944)

\bibitem{Maynard Smith 1982} J. Maynard Smith, Evolution and Theory of the
Games, Cambridge University Press (1982)

\bibitem{Hofbauer2003} J. Hofbauer, K. Sigmund, Bull. Am. Math. Soc. \textbf{%
40}, 479-519 (2003)

\bibitem{Nowak2006} M. Nowak Evolutionary Dynamics: Exploring the Equations
of Life, Belknap Press (2006)

\bibitem{Hauert2005} A. Traulsen, J. C. Claussen, Christoph Hauert, Phys.
Rev. Lett. \textbf{95}, 238701 (2005)

\bibitem{Melbinger2010} A. Melbinger, J. Cremer, E. Frey, Phys. Rev. Lett. 
\textbf{105}, 178101 (2010)

\bibitem{Nash} J. F. Nash Jr., PNAS, 48--49 (1950), Econometrica, 155--162,
(1950), Annals of Mathematics, 286--295 (1951), Econometrica 128-- 140 (1953)

\bibitem{Harsanyi and Selten} J. C. Harsanyi, R. Selten: A General Theory of
Equilibrium Selection in Games. Cambridge, MA: The MIT Press (1988)

\bibitem{Shapley1953} L. S. Shapley, PNAS \textbf{39}, 1095 (1953)

\bibitem{Mertens1981} J. F. Mertens, A. Neyman, Int. J. Game Theory \textbf{%
10}, 53 (1981)

\bibitem{Solan2015} E. Solan, N. Vieille, PNAS \textbf{112}, 13743 (2015)

\bibitem{Arthur1994} W. B. Arthur, Amer. Econ. Rev., \textbf{84}, 406 (1994)

\bibitem{Challet1997} D. Challet, Y.C. Zhang, Physica A \textbf{246}:
407--418 (1997)

\bibitem{Szabo2002} G. Szabo, C. Hauert, Phys. Rev. Lett., \textbf{89}%
:118101 (2002)

\bibitem{Hauert2003} C. Hauert and G. Szabo, Complexity, \textbf{8}%
(4):31--38 (2003)

\bibitem{Hauertetal2002} C. Hauert, S. De Monte, J. Hofbauer, and K.
Sigmund, Science, \textbf{296}:1129--1132 (2002)

\bibitem{Pablo2017} P. A. Valverde, R. da Silva, E. V. Stock, Physica A, 
\textbf{474}, 61-69 (2017)

\bibitem{SilvaIJBC2010} R. da Silva. L. F. Guidi, A. T. Baraviera, Int. J.
Bifurcat. Chaos, \textbf{20}, 1-12 (2010)

\bibitem{SilvaBJP2008} R. da Silva, Braz. J. Phys., \textbf{38}, 74-80 (2008)

\bibitem{SilvaLNE2006} R. da Silva, A.T. Baraviera, S. R. Dahmen, A.L.C.
Bazzan, Lect. Notes Econ. Math., \textbf{584}, \ 221-233 (2006)

\bibitem{SilvaPhysA2006} R. da Silva, A. L. C. Bazzan, A. T. Baraviera, S.
R. Dahmen, Physica. A \textbf{371}, 610-626 (2006)

\bibitem{Harsanyi} J. C. Harsanyi, Management Science, \textbf{14}(3) (1967)

\bibitem{rdasilva2016} R. da Silva, P. Valverde, L. C. Lamb, Commun.
Nonlinear Sci. Numer. Simul. \textbf{36}, 419-430 (2016).
\end{thebibliography}
\end{document}